# THE INTERPLANETARY MAGNETIC FIELD: RADIAL AND LATITUDINAL DEPENDENCES

## Olga V. Khabarova

*Heliophysical laboratory, The Institute of terrestrial magnetism (IZMIRAN), Moscow, Troitsk, Russia*

*E-mail: habarova@izmiran.ru*

Results of the analysis of spacecraft measurements at 1-5.4 AU are presented within the scope of the large-scale interplanetary magnetic field (IMF) structure investigation. The work is focused on revealing of the radial IMF component ($B_r$) variations with heliocentric distance and latitude as seen by Ulysses. It was found out that $|B_r|$ decreases as $\sim r^{-5/3}$ in the ecliptic plane vicinity ($\pm 10°$ of latitude). This is consistent with the previous results obtained on the basis of five spacecraft in-ecliptic measurements (Khabarova, Obridko, 2012). The difference between the experimentally found ($r^{-5/3}$) and commonly used ($r^{-2}$) radial dependence of $B_r$ may lead to mistakes in the IMF recalculations from point to point in the heliosphere. This can be one of the main sources of the "magnetic flux excess" effect, which is exceeding of the distantly measured magnetic flux over the values obtained through the measurements at the Earth orbit. It is shown that the radial IMF component can be considered as independent of heliolatitude in a rough approximation only. More detailed analysis demonstrates an expressed $|B_r|$ (as well as the IMF strength) increase in the latitudinal vicinity of $\pm 30°$ relative to the ecliptic plane. Also, a slight increase of the both parameters is observed in the polar solar wind. The comparison of the $B_r$ distributions confirms that, at the same radial distance, $B_r$ values are higher in low latitudes than in high ones. The analysis of the latitudinal and radial dependences of the $B_r$ distribution's bimodality is performed. The $B_r$ bimodality is more expressed in high latitudes than in the low-latitude solar wind, and it is observed farther in high latitudes. The investigation has not revealed any dependence between $B_r$ and the solar wind speed $V$. Meanwhile, the two-peak distribution of the solar wind speed as measured by Ulysses is a consequence of a strong latitudinal and solar cycle dependence of $V$. It is shown that the solar wind speed in high latitudes (above $\pm 40°$) anti-correlates with a solar activity: $V$ is maximal during the solar cycle minima, and it has a minimum at the maximum of solar activity.

## 1. INTRODUCTION

Investigations of the large-scale structure of the inner heliosphere are actual until nowadays. In spite of the accepted view on the solar wind with a "frozen-in" magnetic field propagating along the Parker spiral, some mismatches between the theory and observations are repeatedly reported [1-3]. Numerous databases of solar wind parameters obtained from the space era beginning allow implementation of more and more detailed multi-spacecraft analysis of the heliospheric plasma properties at different solar cycle phases, heliocentric distances, longitudes and latitudes. As a result, a quantity of the accumulated material transforms into a new quality of understanding of the solar wind processes, but a number of contradictions increases too.

It should be taken into account that Parker's model is stationary (as well as other models using the Parker solution), and some part of deviations from the model may be explained by this fact (history of attempts to improve the original model is presented in the review [4]). Meanwhile, use of a model or hypothesis which stably produces significant mistakes may impact negatively the space weather prognoses quality as well as to be a cause of too slow dynamics in the current scientific area.

Most serious complaints about the quasi-Parker models are mainly lodged by empirics who perform comparative analysis of predictions of the IMF parameters at the Earth's orbit with spacecraft measurements and recalculate the IMF strength from one point to another. The main problem is like this: a solar wind speed (as well as the IMF sign) can be calculated at 1 AU by most models with pretty good accuracy, but the IMF strength and the IMF direction can not be predicted in the same way and with the same accuracy [2, 5, 6]. For instance, the radial IMF component $B_r$ (the RTN coordinate system) theoretically is the most easily predictable IMF parameter, decreasing with distance as $r^{-2}$ in accordance with the Parker's model, but no one of commonly accepted models does provide with adequate $B_r$ prediction at 1 AU [6]. Discrepancies are significant even in solar activity minima, and, during maxima, the correlation between the predicted and calculated $B_r$ values becomes lower any reasonable statistical level. At the same time, semi-empirical models exceed purely theoretical ones by the IMF behavior predictions' quality [7].

Besides, an effect of the "magnetic flux excess" was reported in [8]. The total magnetic flux in the heliosphere $F_s$ can be calculated as:

$$F_s = 4\pi |B_r| r^2 \qquad (1)$$

According to the Parker's theory, $F_s$ should be a constant everywhere in the heliosphere under at least 27 days (one Carrington rotation period) averaging, but in fact there is a difference between $F_s$ from distant spacecraft and near-Earth measurements [8]. The difference was found to be increasing with distance. It becomes so expressed at $r > 2.5$ AU that it can not be ignored.

Many works were dedicated to investigations of the "flux excess" effect, and several explanations were suggested (from the kinematic effect to wrong averaging methods) [9-11].

In the original paper [8], the mean of the module of the radial IMF component ($<|B_r|>$) was used, but authors of some other papers prefer to use a module of the mean $|<B_r>|$ [9-11]. It should be noted that $|<B_r>|$ and $<|B_r|>$ are not identical as $B_r$ ranges from negative to positive values. Sometimes it is supposed that there is no difference between $|<B_r>|$ and $<|B_r|>$ for a rather large temporal averaging interval, sometimes the opposite opinion prevails [8-12].

Absence of clear understanding of physical causes of the effect leads to specific correction factors application, special methods of data processing and other artificial techniques use. At the same time, correctness of the formula (1) is supposed to be beyond doubt. Let's consider here a hypothesis that the problem is not in some unclear physical effects, but in the fact that $B_r$ depends on the heliocentric distance not as $r^{-2}$. This question was raised by the author in [13], and it will be discussed among other questions in the current paper in details.

The point-to-point IMF recalculations in the heliosphere and $F_s$ conservation are based on the Parker's theory invariant $B_r \cdot r^2$. Meanwhile, according to the recent multi-spacecraft data analysis, $B_r \cdot r^2$ is not conserved in the inner heliosphere [13]. The Helios 2, IMP8, Pioneer Venus Orbiter, and Voyager 1 spacecraft IMF measurements from 0.29AU to 5.0 AU in the increasing phase of the solar cycle (from 1976 to 1979) were considered in [13]. Results of the Parker's model (recalculations of the source surface magnetic field along the Parker spiral as $r^{-2}$), as well as the pure radial expansion model results (when the magnetic field decreases as $r^{-2}$, but not along the spiral) are given in Fig. 1a, which represents the merged figures 1-2 [13].

The observations show that the radial IMF component decreases with distance not as $r^{-2}$, but as $r^{-5/3}$. At the same time, the tangential component behavior was found to be corresponding to the expected law ($r^{-1}$), and the IMF strength $B \propto r^{-1.4}$. The difference between the observed and calculated $B_r$ values is most expressed at small heliocentric distances, where the observed field may several times exceed the values calculated through the models using the Parker solution. At $r > 5$ AU, the difference is not so significant.

The model's quality can be improved through the application of some corrections, for example, by multiplying the obtained values by some constant or summarizing them with some additional field. In the result, the both model curves seen in Fig. 1a below the experimental one will be shifted up and coincide with observations better. These methods are usually used for best correspondence of calculations with 1 AU observations. The fitting methods are typically performed just for two points: "the source surface" – "the Earth's orbit". Unfortunately, as it was mentioned above, the acceptable correspondence is not reached even at 1 AU, and discrepancies get more significant at other distances.

In addition, another consequence of insufficient understanding of the large-scale IMF behavior was demonstrated in [13]: there is an effect of unexpected vanishing of the $B_r$ distribution's bimodality with distance. At the Earth's orbit, the $B_r$ distribution has a well-known two-humped (bimodal) view because of the expressed sector structure of the IMF in the Earth's vicinity. As a result, the $B_r$ histogram consists of two quasi-normal overlapping distributions corresponding to the IMF measurements in positive and negative sectors. Before [13], it was commonly supposed that such a picture should be observed at least up to the first turn of the Parker spiral, when the spiral gets perpendicular to the sunward direction (which occurs farther 5 AU at any solar wind speeds). Meanwhile, the experimental data by Helios 2, IMP8, Pioneer Venus Orbiter, Voyager 1, and Ulysses do not confirm that. In fact, bimodality vanishes with distance very fast: it is clear at 0.7-1.0 AU, it is seen at 2-3 AU worse, and it fully disappears at 3-4 AU.

Most possible, this is a consequence of a radial increase of the solar wind turbulence which results in disappearance of the clear sector structure much closer to the Sun than it was supposed earlier. Indeed, one of the IMF modeling problems is mismatching of predicted and observed localization and inclination of the heliospheric current sheet (HCS). Spatial and temporal parameters of this greatest structure in the heliosphere determine the whole picture of the IMF, but the HCS features are known poorly. The HCS Parker's angle at different AU, large-scale HCS twisting as well as the south-north displacement are widely discussed questions, being of great importance for the IMF properties revealing [14-16]. Unfortunately, as shown in [17] through the comparison of observations and calculations according to the Stanford source surface magnetic field model, predictions of the HCS position meet with serious problems. The difference between the calculated HCS azimuth angle and experimentally found values sometimes reaches 25°. The best results can be obtained by use not MHD-, but semi-empirical models [18-19]. All discussed facts demonstrate insufficient understanding of processes of the solar wind expansion into the heliosphere.

Therefore the IMF behavior in the inner heliosphere seriously differs from predictions of quasi-Parker models even under a rough approach, and demands further studies. On the one hand, Parker's model is very attractive by its simplicity and a possibility of fast solar wind parameters point-to-point recalculations through elementary formulas. On the other hand, there is obvious necessity to understand why those formulas work well for plasma parameters, but do not work for the IMF, and what the true IMF radial decrease law is in the inner heliosphere.

The empirical investigations of the large-scale IMF picture in light of the discussed problems and already obtained results will be continued in the current paper. The main idea is as follow: the revealed deviations of the experimentally observed IMF parameters from the predicted ones are mainly not caused by non-stationary effects (such as CMEs), but are related to misused IMF recalculation according to $r^{-2}$ law. Also, a-priori believing the IMF to be completely "frozen in" to the plasma may lead to the observed discrepancies.

A new model development is a future task. Meanwhile, there are several key dependencies which may be empirically found right now. For example: What is the law of the IMF decrease in the inner heliosphere? What is the main cause of the "flux excess" effect? Whether the IMF depends on solar wind speed and heliolatitude? These (as well as related) questions will be discussed below.

# 2. PECULIARITIES OF THE INTERPLANETARY MAGNETIC FIELD RADIAL AND LATITUDINAL DEPENDENCIES

## 2.1. The radial IMF component changes with distance. The "magnetic flux excess" problem.

Let's check the results obtained in [13] by the calculation of the $B_r$ curve slope through the alternative spacecraft data analysis. The best candidate for this purpose is the Ulysses spacecraft, which allows consideration both radial and latitudinal dependencies because of its unique orbit, which was nearly perpendicular to the ecliptic plane. Hourly Ulysses data were used in this work for 25.10.1990–30.09.2009 (see http://cdaweb.gsfc.nasa.gov/).

For the adequate comparison, let's select the data for the near-ecliptic (±10° latitudinal vicinity around the ecliptic plane) Ulysses passages. Then, the "|<$B_r$>| or <|$B_r$|>" problem arises (see [10]). Most reasonable and reconciling approach to the problem's solving was demonstrated in [11]. Meanwhile, in the author's subjective opinion, first-step averaging of a bimodally distributed parameter is unreasonable. For instance, averaging of any sinusoid gives zero. The module of the pre-averaged $B_r$ (|<$B_r$>|) gives an analogical result. It is not zero for short-time intervals of averaging. But the longer time interval is considered, the result is closer to zero (which is the mean of the symmetrically distributed bimodal parameter $B_r$). As a result, the longer time interval, the lesser the "flux excess", but the physical sense of the results of such averaging is as disputable as a sense of the averaged sinusoid.

To avoid the "averaging mistake", let's consider here the radial IMF component module ($|B_r|$) without averaging used in [13]. The result is shown in Fig. 1b, analogues to Fig. 1a. The white approximation curve has a slope of -1.614, which is very close to -5/3 found in [13], hence $B_r \propto \cdot r^{-1.6}$. It is interesting that the used Ulysses data covered much more extended temporal interval in comparison with the data taken for the previous analysis in [13]. Similarity of the results means stable deviations of the IMF behavior from theoretical assumptions.

Let's turn now to the physical meaning of the factor $x$ multiplying $r$ in the approximation equation $|B_r| = x \cdot R^{-y}$.

In the paper [13], $x=3.8$ (see Fig. 1a), but $x=2.4$ in Fig. 1b. It is known that after multiplying of a function by a positive number, the graph stretches up. In our case, the essential difference will be seen in rising of the graph's part corresponding to small distances from the Sun. According to a method of dimensions, $x$ from $x \cdot R^{-y}$ can be represented as $x=B_0$, where $B_0$ is some reference field, and $R=r/r_0$ ($r$ is a heliocentric distance, a variable; $r_0$ is a distance from the Sun to the point, where $|B_r| = B_0$). Therefore, the radial IMF component is:

$$|B_r| = B_o \left( \frac{r}{r_0} \right)^{-5/3} \qquad (2)$$

In [13] (see Fig. 1a), $B_0 = B_{IAU}$ at $r_0 = 1$ AU as obtained on the basis of measurements, starting from the heliocentric distance of 0.29 AU. In the case of the Ulysses database use (Fig. 1b), the IMF was measured farther 1 AU, and $B_0$ had a smaller value.

It is important to note that $B_0$ varies with time and solar cycle. Its measured value is also influenced by the spacecraft magnetometer's characteristics. As an example of the IMF temporal variations, one can see the features of the IMF changing with solar cycle at 1 AU in the first figure of the paper [20]. According to $B_0$ changes, the entire experimental curve correspondingly shifts up and down, but the slope most possibly remains the same.

Let's take as a basis that the $B_r \propto \cdot r^{-2}$ statement is true just for the first-approach estimations. Deviations of $y$ from 2 may lead to serious mistakes of $B_r$ point-to-point recalculations. The "flux excess" effect is just a confirmation of this statement. Theoretically, $F_s$ should be an invariant over the entire heliosphere, at any distances. As it was mentioned above, a consistent increase of $F_s$ with heliocentric distance is observed: distantly measured $F_s$ increasingly differs from $F_s$ calculated on the basis of measurements at the Earth's orbit. This deviation from the theory may be neglected (as many other discrepancies are ignored) if one's interest is just in the estimation of an order of values, but, undoubtedly, such intriguing inconsistency is worthy to be a subject of keen interest.

The assurance that $B_r$ decreases as $r^{-2}$ leads to the situation, when among different assumptions on the nature of the effect, a hypothesis about the inapplicability of the theory due to non-ideality of space plasma was not seriously considered. Meanwhile, one can see that the observed effect may be explained easily. Let's calculate the difference between the magnetic flux $F_s(r)$ at some heliocentric distance $r$ and $F_{s\_1AU}$ at 1 AU, taking into account (2):

$$\Delta F_S = F_S(r) - F_{S\_1AU} = 4\pi\left(|B_r|\,r^2 - B_{1AU}[1AU]^2\right) = 4\pi\left(B_{1AU}\left(\frac{r}{[1AU]}\right)^{-5/3}r^2 - B_{1AU}[1AU]^2\right) =$$

$$= 4\pi\frac{B_{1AU}}{[1AU]^{-5/3}}\left(r^{2-5/3} - [1AU]^{2-5/3}\right) \qquad (3)$$

In some papers (see, for example, [11, 12, 20]) factor of $2\pi$ is used instead of $4\pi$, as half of the flux is directed away from the Sun and half is sunward, but it does not change the matter of the effect. This dependence is graphically represented in Fig. 2 as a black curve. Points are $\Delta F_s$ taken from [8]. In [8] $\Delta F_s$ was calculated on the basis of several spacecraft data and averaged by 0.1 AU (the corresponding deviations can be found in figure 5 from [8]). The standard assumption on $|B_r|$ changing with distance as $r^{-2}$ was used in [8] for the $\Delta F_s$ calculations.

Fig. 2 demonstrates rather good correspondence of the data [8] and the curve (3). For example, theoretical curves in [9] calculated in assumption of kinematic effects had upward trend. It is remarkable that the calculations according to (3) do not need anything but heliocentric distance. Moreover, the suggested approach explains easily the mysterious "separating point" of $\Delta F_s = 0$ ($\Delta F_s$ is negative before and positive after it). It is easy to see from (3), that this point is 1 AU at the axis of abscises (it is shown by light-grey lines in Fig. 2). This is a natural consequence of the fact that $F_s$ in [8,9] was calculated on the basis of practically same data as taken in [13] for derivation of the $|B_r| = B_{1AU}\left(\frac{r}{[1AU]}\right)^{-5/3}$ dependence. Therefore, $B_0$ in [8, 9] and in [13] coincides: $B_0 = B_{1AU}$. Obviously, this point may slightly shift under changing of the selected database, as $B_0$ changes.

Certainly, the effects discussed in [9-12, 20] also exist and amplify the revealed trend. Among other things, the flux excess effect can be additionally determined by the latitudinal dependence of $B_r$, which will be discussed below. It should be pointed out that the difference between the calculated and observed flux (depending on distance as $R^{-y}$) appears at any $y \neq 2$.

Hence, there is necessity of further investigation of the $B_r$ behavior in the inner heliosphere as seen by different spacecraft at different phases of solar cycle.

## 2.2. Latitudinal and solar cycle dependences of $B_r$

To answer the question about the dependency or independency of $B_r$ on latitude, let's look at the whole picture of the solar wind parameters' changes provided by Ulysses. Hourly data for all the period of measurements are shown in Fig. 3. The spacecraft's trajectory is presented in the upper panel (a), where latitude is shown by white and the heliocentric distance $r$ – by black color. Referring to it, three areas of the increased amplitudes and disturbances of all parameters are seen in other panels. They correspond to $r \leq 2$ AU and the fast heliolatitude change. Maximum of the parameters' changes accrues to the spacecraft's crossings of the ecliptic plane.

Guided by morphological data analysis, investigators made conclusions that the solar wind speed decreases and density increases ± 30° around the ecliptic plane (see, for example, [20, 21], http://ulysses.jpl.nasa.gov/science/mission_primary.html and http://ulysses.jpl.nasa.gov/2005-Proposal/UlsProp05.pdf). At the same time, it is accepted that the radial component of the IMF has no latitudinal dependency at any certain heliocentric distance. This statement is used as a basis of many works ([8-12, 20-23]).

The recent paper [23] gives typical views on this topic: $B_r$ does not depend on latitude, as "the magnetic flux density (referred to 1 AU) tends to be uniform, at least in the fast, polar solar wind <...> the magnetic flux density measured at a single point is a representative sample of the absolute value of the magnetic flux density everywhere in the heliosphere." (page 2 [23]). As was mentioned above, such an approach gives gratifying results at the first approximation, but more detailed analysis can give a key to the explanation of many inconsistencies between the theory and observations.

Unbiased look at the picture of the radial (b), tangential (c) IMF components and the IMF strength (d) in Fig. 3 may put in doubt the statement of latitudinal independency of $B_r$, because the IMF (as well as all its component's) growth, occurring simultaneously with the solar wind speed decrease and density increase, is rather obvious. Meanwhile, taking into account the double $B_r$ dependence (on both latitude and distance), an additional statistical analysis must be performed. Let's separate variables and investigate how the radial IMF component varies with latitude and heliocentric distance. Then, a least-square 3-D surface «$B_r$–latitude–distance» can be plotted (see Fig. 4).

The radial IMF component in the subspace «$B_r$–heliolatitude» (Fig. 4a) has two trends: $B_r$ increases toward the ecliptic plane, and there is some less-expressed $B_r$ enhancement in the polar latitudes. Fig. 4b represents $B_r$ changing with distance. The |$B_r$| radial decrease was partially studied in the Section 2.1. Combined picture of these two dependencies is shown in Fig. 4c as a 3-D surface.

The same kind of surface for the IMF strength $B$ is given for the comparison (Fig. 4d). Both panels demonstrate very similar behavior of $B_r$ and $B$: the magnetic field decreases with distance, and it has maximum at the ecliptic plane (this increase is most expressed at small distances from the Sun). Slight increase of $B_r$ and $B$ in polar regions is an interesting feature which may be related to peculiarities of the solar magnetic field generation (dynamo waves effect) as was predicted in [24].

The found effect of the IMF strength increase at the ecliptic plane may be proved by another way. Let's analyze Ulysses data at the three selected distance ranges from the Sun (1-2 AU, 2-3 AU, and 3-4 AU) and separate them by latitude (above and below 40°). The radial component distribution's view at different latitudes and heliocentric distances is shown in Fig. 5ab. One can see that at the same distance $r$ from the Sun, the high-latitude $B_r$ values are always smaller in comparison with low-latitudes values. This is a plain evidence of the found effect. Therefore, the radial IMF component can not be considered as independent of heliolatitude because of its pronounced increase it the ecliptic plane vicinity, especially at small heliocentric distances.

Continuing the $B_r$ histogram analysis, it is necessary to mention the $B_r$ "bimodality effect" (see Introduction) which became an object of special interest during the last years. It is known that at the Earth's orbit, the horizontal IMF components (the GSE coordinate system) or radial and tangential components (RTN, in the ecliptic plane) have a two-humped view. Initially, the histograms view variations with solar cycle was investigated in [25]. Then, the work on the histograms' change with distance followed [26] (this preprint's results were partially published in [13]). The question about the dependence of the magnetic flux density ($B_r \cdot r^2$) histogram shape of the solar wind flow type (high-speed, low-speed and CME) as well as of a solar cycle was discussed in [23].

It is shown (Fig. 5) that the effect of fast $B_r$ histogram's bimodality disappearance found in [13, 26] for the IMF in-ecliptic measurements looks significantly smoothed at high latitudes. Two peaks transformation into one peak with increasing $r$ is clearly seen in Fig. 5b (low latitudes), but it is not so expressed in Fig. 5a (high latitudes). Obviously, at high latitudes, the bimodality disappears farther from the Sun.

It would be interesting to trace the $B_r$ histogram's shape change at high and low latitudes in different solar cycle phases. One can analyze the horizontal IMF (GSE) components' histograms at 1 AU in the ecliptic plane through OMNIbase data, collected for a long time by many spacecraft (see. Fig. 6ab). The histograms become wider at solar activity maxima, their peaks get down, but the bimodality effect does not disappear. It is useful to note for the further comparisons that the $B_x$ IMF component in GSE coordinate system equals to $-B_r$ in the RTN coordinate system.

Relation of the $B_r$ histogram shape in high latitudes to the solar activity cycle is shown in Fig. 6cd, which represents parts of Fig. 5a selected by cycle phase. Ulysses observations at the latitudes above 40° covered two minima and one maximum of sunspot numbers. At solar activity minima, the histograms' bimodality is expressed more clearly than during the solar maximum. Prevailing of one or another histogram's hump in Fig. 6cd is related to statistical prevailing of positive/negative active region on the Sun. As a whole, the bimodality effect disappears neither in minimum, nor in maximum of solar activity at high latitudes.

During the solar activity maximum, the histograms' spreading is observed at all distances from the Sun. This effect is known in the ecliptic plane, 1 AU (see. Fig. 6ab), where it usually is explained by the impact of CMEs, which occurs most frequently during the solar activity maxima. Hence, the observed high-latitude histograms' spreading is an indirect sign that CMEs fill the significant part of the inner heliosphere.

Temporal (solar cycle) changes of the interval between the positive/negative histogram peaks as seen in Fig. 6ab were investigated in [25], where the IMF strength along the Parker spiral $B_L$ was calculated on the basis of OMNI data. It was found that the variations of the distance $\Delta B_L$ from one peak to another have the same tendency as was demonstrated in Fig. 6: $\Delta B_L$ is maximal in maxima of solar activity and minimal in the solar activity minima. One can see that the distance $\Delta$ between the humps of a histogram correlates with the peaks' height. If the histogram spreads, the peaks' height drops. This helps to fill the gap in our knowledge of the IMF behavior at high latitudes. There is no sufficient information from Ulysses on the IMF there during solar maxima, but as the solar cycle dependencies of the $B_r$ bimodality are same at any latitudes, it is possible to expect that $\Delta B_r$ varies with cycle in the latitudes above 40° in the same way as it is seen in low latitudes.

## 2.3. Whether the radial IMF component depends on the solar wind speed? The solar wind speed changes with latitude and solar cycle.

It is remarkable that Fig. 5 and Fig. 6 are consistent with some results of [23]. The magnetic flux density ($B_r \cdot r^2$) bimodality is expressed in the "fast" solar wind rather than in the "slow" solar wind as follows from figure 1 in [23]. The solar wind speed $V$ measured by Ulysses has a two-peaks distribution, so separation of the "fast/slow" solar wind was made in [23] according to this fact. Having in mind that $V$ has only one-peak distribution at the Earth's orbit, let's puzzle out why Ulysses has seen those two peaks. What does the "fast" or "slow" solar wind mean when we use Ulysses measurements?

$V$ is believed to be approximately independent of latitude and longitude in [23]. The solar wind types are divided according to the Ulysses-measured $V$ distribution peaks: the "slow wind" has velocities $V < 400$ km/s and the "fast wind" flows faster than 600 km/s independently of latitude or heliocentric distance. Meanwhile, more detailed analysis does not confirm that the $B_r$ (and, consequently, the magnetic flux density) is determined anyhow by $V$. A thesis about the latitudinal independency of $V$ is not confirmed either. As one can see below, the found in [23] dependence of $B_r$ on the fast/slow solar wind, in fact, is a latitudinal dependence.

The solar wind speed dependencies are shown in Fig. 7-9. Fig. 7a is illustrative of a scatter of points in the "radial IMF component – solar wind speed" subspace, where two clouds of points are seen. There is no $B_r$ dependency of $V$ inside each cloud. As one can see below, these two clouds correspond to two humps of the $V$ distribution. To answer the question about nature of the mentioned bimodal $V$ distribution, it is necessary to plot the Ulysses "latitude-distance" curve (Fig. 7b), as well as to reveal the "speed-latitude" and "speed-distance" dependencies (Fig. 8).

As seen in Fig. 8, the $V$ distribution's bimodality (Fig. 8a) is a consequence of the solar wind complex latitudinal (Fig. 8b) and radial (Fig. 8c) dependencies. Fig. 8b resembles a flying eagle with two wings and two legs (one leg, corresponding to negative latitudes, is more expressed). The radial solar wind speed dependence is shown in Fig. 8c. It is also degenerate, as there are both lower and upper branch of the curve. Fig. 8d was built in the same manner as Fig. 4cd. One can see there "wings" of Fig. 8b and some $V$ increase at 2-3 AU of Fig. 8c, as well as well-known strong $V$ decrease in the area of zero latitude (the ecliptic plane).

Nature of the $V$-"wings" and "legs" in Fig. 8b is related with both the Ulysses' trajectory features and the solar cycle. "Legs" represent $V$ in high latitudes in the solar activity maximum, and "wings" correspond to $V$ measurements in high latitudes during solar activity minima. This follows from Fig. 9 where $V$ was plotted separately for the latitudes above ±40° (Fig. 9a) and for the latitudes ±10° around the ecliptic plane (Fig. 9b) in comparison with the solar cycle. Fig. 9a shows that the solar wind speed was minimal in the maximum of solar activity and maximal during the solar activity minima.

Therefore, the high-latitude solar wind is fast (as it commonly believed) at solar activity minima only. Comparison of Fig. 9a and Fig. 9b allows find that $V$ in high latitudes during the solar maximum has values close to $V$ in low latitudes. It is interesting that variations of the near-ecliptic solar wind speed have ~ 2 years outstripping shift regarding the solar cycle maxima/minima. There is a peak of 2002-2005, known by its anomalies. As it was shown in [27], the large-scale magnetic field of the Sun unexpectedly dominated during that period, and this is seen in the solar wind too. Anyway, the solar wind speed relation with a soar cycle is obvious just in high latitudes. Near the ecliptic plane, it is still doubtful.

Summarizing, one can see that a bright solar cycle dependence of $V$ is the main source of the $V$ distribution bimodality. The central part of Fig. 8b (the "eagle's body") is $V$ in low latitudes; the high-latitudinal solar wind in solar activity minima forms two upper branches ("wings" of $V > 700$ km/s). Besides, the high-latitude solar wind in the maximum of solar activity inputs $V$ values approximately from 400 km/s to 600 km/s (see the "legs" in Fig. 8b). Weakness of this branch is explained merely by insufficiency of measurements in a solar maximum (Ulysses provided high-latitude measurements for two minima and for one maximum of solar activity only).

A hypothetical continuation of the Ulysses measurements would lead to enhancement of the "legs", and the whole picture of the solar wind speed latitudinal dependence would be completed. The "wings" in Fig. 8b form the 650-850 km/s peak of the $V$ distribution (Fig. 8a). The "legs" and the "eagle's body" give the second peak of the $V$ distribution with values of 250-550 km/s. Thus, the $V$ distribution bimodality is determined by changing of Ulysses' latitude and solar cycle.

This effect may lead to some misunderstandings. For example, the authors of [23], in fact, did not study the magnetic field of the "fast" solar wind, but revealed the high-latitude solar wind properties in solar activity minima. Sure, this fact does not diminish the [23] results, but demands a right approach to them. It is obvious from Fig. 8b that the lower threshold of the "fast" solar wind (600 km/s), as selected in [23], corresponds to a super-fast stream in low latitudes, but in high latitudes such a stream would be super-slow. This also should be taken into account in further investigations. Most reasonable "fast/slow wind" separation has been already made in [28], where the threshold of 450 km/s was used, and its physical causality was confirmed.

Regarding the radial dependence of $V$, the lower branch in Fig. 8c, rising with distance according to theoretical expectations, mainly belongs to Ulysses measurements in low latitudes, and the branch, decreasing after 3 AU, primarily corresponds to measurements in high latitudes. The "primarily" word is used here because there is also the solar cycle $V$ dependence. Furthermore, Ulysses registered fast streams in high latitudes even during the solar activity maximum (mainly in 2001-2002). Meanwhile, as a whole, the solar wind speed decreases with distance in high latitudes at solar activity minima, and $V$ radially growth in low latitudes independently of solar activity (see Fig. 10). Two solar activity minima data were selected for high latitudes (Fig. 10a), and all available data for near-ecliptic measurements (±10° of heliolatitude) were taken for Fig. 10b.

A "noisy" part of the data in Fig. 10b is a consequence of the Ulysses orbital rotation, as a spacecraft has a minimal velocity in its apogee, so the $r$ interval of > 5 AU contains more points and the curve is noised by non-stationary effects such as CME. The same effect is seen in Fig. 1b and Fig. 4b.

According to Fig. 7-10, the solar wind speed depends on heliolatitude and solar cycle phase in a high degree. At the same time, the radial IMF component does not depend on solar wind speed.

## 3. CONCLUSIONS AND DISCUSSION

Several problems important for understanding of the large-scale picture of the magnetic field in the inner heliosphere were discussed in the paper:

1. What is the law of the radial IMF component ($B_r$) decrease with heliocentric distance?

2. The "magnetic flux excess" is radial increase of the open solar flux as calculated from distant spacecraft measurements in comparison with 1 AU flux. What is the cause of the "magnetic flux excess"?

3. Whether $B_r$ depends on heliolatitude?

4. How does the $B_r$ distribution's bimodality vary with latitude and solar cycle?

5. Is there the $B_r$ dependence of the solar wind speed $V$?

6. How does $V$ depend on heliolatitude, heliocentric distance and solar activity?

The results of the Ulysses and OMNI data analysis show that listed above problems are related. For example, confirmed in this work deviation of the $B_r(r)$ law from the classical dependence used in Parker-like models is one of the main causes of the "magnetic flux excess". The latitudinal $B_r$ dependence also results in this effect. At the same time, the dependence of the magnetic flux density on the solar wind speed reported in [23], in fact, is the $B_r$ dependence on latitude and solar cycle.

Correspondingly to the listed above questions 1-6, it was found out that:

1. *The radial IMF component $B_r$ decreases as $r^{-5/3}$, but not as $r^{-2}$.*

This conclusion was made on the basis of the Ulysses data for the entire period of measurements. The $B_r$ module approximation was used to avoid the "module of the mean or mean of the module" problem (see Section 2.1). The same result was obtained in [13] from analysis of the five spacecraft data.

The phenomenon's nature may be not only in the fact that Parker's model is stationary, but also in poor applicability of the "frozen-in" magnetic field assumption to the non-ideal space plasma conditions. Indeed, the "frozen-in" IMF conditions' break occurs in the solar wind very often, for example, in some vicinity of current sheets. As was shown [13], zero IMF lines corresponding to current sheets are observed in the solar wind inside the IMF sectors more frequently than it was supposed earlier (zero lines were expected to be observed mainly at the heliospheric current sheet).

A magnetic reconnection recurrently occurs at the large-scale heliospheric current sheet as well as at smaller-scale current sheets during the solar wind expansion. As a result, current sheets are subjects of a multiplication (bifurcation) process. A significant part of the heliosphere is filled with secondary current sheets and other products of the magnetic reconnection in some vicinity of the current sheets. Under averaging, it looks as a radial increase of turbulence and intermittency of the solar wind plasma, and, finally, as a break of the expected IMF radial dependence law. It is worthy to remark that the solar wind plasma obeys the Parker's theory much better than the IMF does. This is also a confirmation that the IMF is not fully frozen into the solar wind plasma.

2. *The «magnetic flux excess» ($F_s$) is mainly a consequence of the conclusion 1.*

Accepting that the $B_r$ decreases as $r^{-5/3}$, it is easy to explain the experimentally calculated values of the excess $\Delta F_s$ – i.e. the difference between $F_s$ obtained through distant spacecraft data and the measurements at the Earth's orbit. Obviously, any deviation of the real law of the $B_r$ radial decrease from the theoretically expected leads to unavoidable dissimilarities at the point-to-point recalculations through $B_r\,r^2$ formula. Thus, an experimental study of the radial

IMF dependence observed by different spacecraft seems to be a perspective way of future investigations. The latitudinal IMF dependence as well as effects discussed in [11, 12, 20] contribute to the "flux excess" effect.

*3. The radial IMF component depends on heliolatitude.*

$B_r$ as well as the magnetic flux can be considered as independent of heliolatitude just in a rough approximation. More detailed investigations show the radial IMF component and the IMF strength increase toward the ecliptic plane. Additionally, some IMF enhancement is observed in the polar solar wind. The result is checked by different methods, including the analysis of the $B_r$ histograms at different heliocentric distances.

Most probably, the discussed underestimation of the IMF latitudinal dependence appears from history of development of views on the large-scale solar magnetic field. Before the Ulysses mission, there was a dominating opinion that the magnetic field of the Sun should be similar to the Earth's magnetic field and it was very close to classic magnetic dipole. The polar magnetic field strength was expected to be twice of equatorial. The Ulysses data did not confirm that. So, in the midst of the expected difference between the polar and low-latitude IMF, the picture observed by Ulysses looked as any absence of the latitudinal dependence of the IMF.

It is necessary to remark that the statement of the $B_r$ latitudinal independency seems very strange from the view of investigators of the solar processes. Zonality and difference of the solar magnetic field properties at low and high latitudes are obvious and proved by long time observations. It seems to be extremely unlikely that all observed differences take place only at distances below ten solar radii and then they disappear (the open magnetic flux uniformity demands such an assumption).

Discarding of the $B_r$ and $B$ increase in low heliolatitudes inevitably reduces a quality of even very competent models (such as [29] and [30]), as they are based on slightly simplified views on the large-scale IMF picture in the inner heliosphere, assuming a constant $B_r$ at any distance with a sharp [29] or more gradual [30] change of the $B_r$ sign at the ecliptic plane.

4. a) *The $B_r$ histogram's bimodality is expressed in high heliolatitudes (above 40°) rather than in low latitudes. It can be observed in high latitudes at those heliocentric distances, where it already vanished near the ecliptic plane.*

This fact may bear evidence of radial increasing of turbulence and intermittency in the solar wind due to mentioned above processes in current sheets (and, most notably, in the heliospheric current sheet). Indeed, unimodality of the $B_r$ histogram means the absence of any clear sector structure. Most possibly, mixing of structures occurs in low latitudes at 3-4 AU, but in high latitudes the solar wind remains well-structured at the same distances.

*b) In high heliolatitudes, the $B_r$ histogram's view has the same solar cycle dependence as in low latitudes: in the solar maximum the distribution spreads, and its peaks descend.*

This means that in solar activity maximum both $B_r$ and the $B_r$ internal scatter increase. Most possibly, this is a consequence of the CME impact on the high-latitude solar wind during solar maxima, which confirms the observers' conclusion that CMEs fill a significant part of the inner heliosphere [31].

*5. $B_r$ is independent of the solar wind speed.*

*6. a) The solar wind speed significantly depends on heliolatitude. In high latitudes, it strongly depends on solar activity.*

The solar wind speed increase in high latitudes (in comparison with its values near the ecliptic plane) has been known since the first Ulysses flyby. After that, the solar wind above 40° has been believed to be fast.

The current investigation revealed that aforesaid is true only for minima of solar activity. During a solar activity maximum, the high-latitude solar wind speed decreases to values, typical for low heliolatitudes. The difference between the yearly mean high-latitude $V$ in maximum and minimum is 200-300 km/s.

The latitudinal dependence of $V$ has four branches resembling "an eagle with outspread wings". The solar wind flows above ±40° form the high-speed "wings" during solar activity minima. At the solar activity maximum, lower high-latitude branches are formed (the "legs" with $V \sim$ 270-500 km/s). The low-latitude solar wind speed is characterized by values of 300-550 km/s well-known through the data of in-ecliptic spacecraft.

*b) The solar wind speed depends on distance differently at high and low latitudes.*
There are two branches of the radial $V$ dependence. The lower one is mainly formed by $V$ measurements in low latitudes, when $V$ expectedly growths with distance. This branch also contains a significant part of data obtained in high latitudes in solar activity maximum.

The upper branch mainly corresponds to high-latitudinal $V$ measurements in solar activity minima. It looks like an arch having maximum at ~ 2-3 AU. As a whole, $V$ in high latitudes decreases with heliocentric distance. Further investigations must be carried on to find why the coronal hole's plasma expands with decreasing speed.

Therefore, using the solar wind speed data by Ulysses, it is necessary to take into account the follows:

- the high-latitude solar wind is not permanently fast;

- the *V* histogram's bimodality is a consequence of latitudinal and solar cycle dependencies of the solar wind speed;

- the solar wind speed increases with distance in low latitudes as well as in high latitudes during solar activity maximum, but the high-latitude *V* radially decreases in solar activity minima.

All found peculiarities of the solar wind plasma propagation may be used in advanced models such as [32].

All the discussed effects together demonstrate that the observed solar magnetic field and plasma's features are clearly seen in the solar wind at rather far distances from the Sun, farther the Earth's orbit. The Ulysses measurements have revealed both solar wind zoning and distinctions of the solar wind propagation in different phases of the solar cycle.

Keeping in mind all above said, one can see a substantial input of high-latitude missions into development of views on the magnetic field in the heliosphere. The Ulysses mission provides nutriments for long-time investigations. Meanwhile, statistical data insufficiency does not allow detailed analysis of the solar cycle dependences of solar wind parameters, and, in some measure, there is no enough information on the radial IMF variation in heliosphere. Many hopes are anchored now on the future Russian Interheliozond mission. Meantime, the obtained results will be validated and added through the analysis of different missions' data available by now.

The Ulysses data are taken from the Coordinated Data Analysis (Workshop) Web-site: http://cdaweb.gsfc.nasa.gov/ (the magnetic field and plasma data were provided by Prof. A. Balogh and Dr. John L. Phillips, Imperial College, London, UK).
OMNI data are obtained from the Goddard Space Flight Center OMNIweb plus web-site: *http://omniweb.gsfc.nasa.gov* .
The author cordially thanks Prof. Vladimir Obridko and Dr. Kirill Kuzanyan for fruitful discussions.
This study was supported by the RFBR grants №11-02-00259 и №13-02-92613.

*IZMIRAN, Troitsk, Moscow, Russia 124190*

*Tel: +74958515567; Fax: +7 (495) 851-01-24; e-mail: habarova@izmiran.ru*

FIGURES

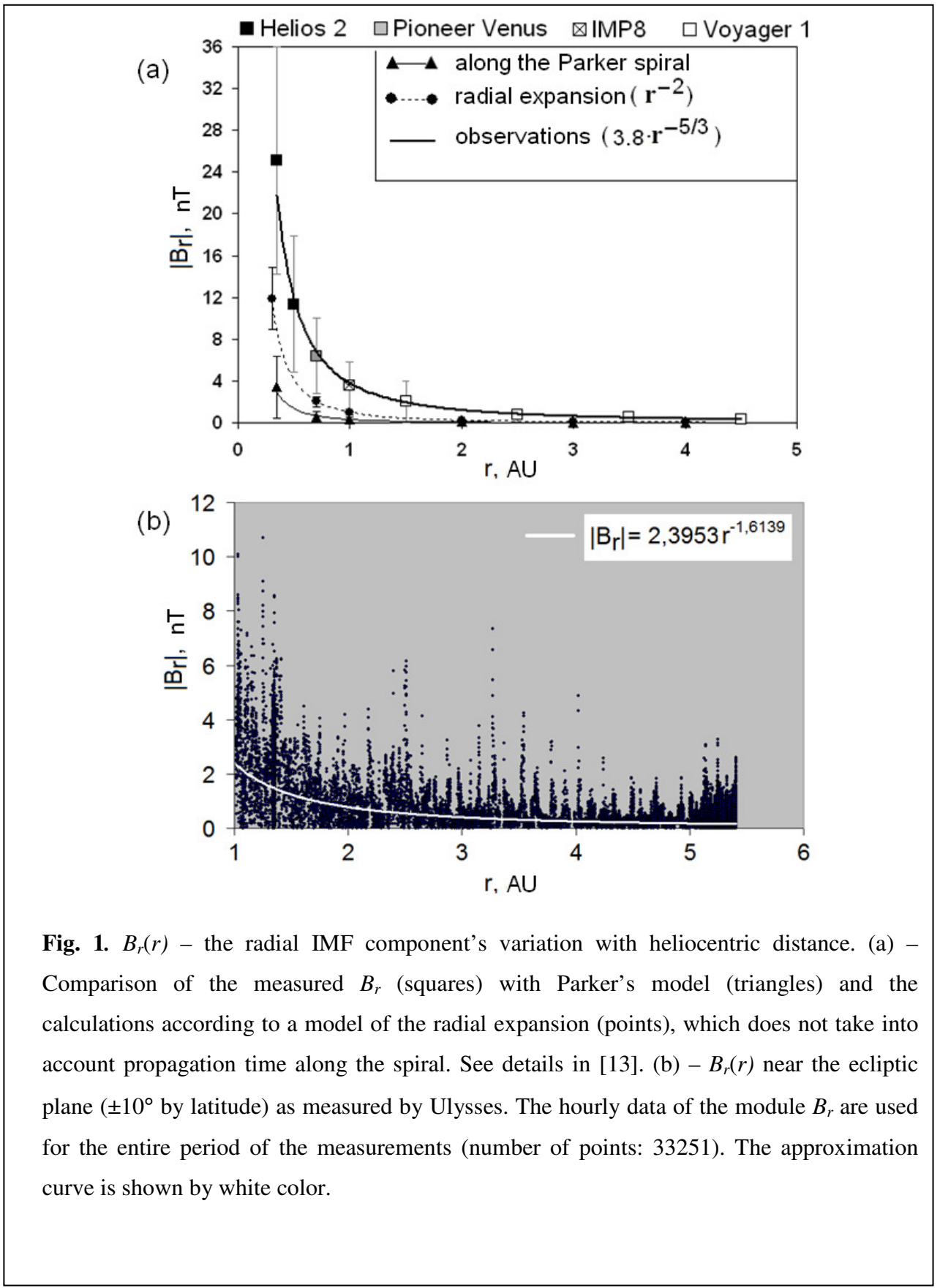

**Fig. 1.** $B_r(r)$ – the radial IMF component's variation with heliocentric distance. (a) – Comparison of the measured $B_r$ (squares) with Parker's model (triangles) and the calculations according to a model of the radial expansion (points), which does not take into account propagation time along the spiral. See details in [13]. (b) – $B_r(r)$ near the ecliptic plane (±10° by latitude) as measured by Ulysses. The hourly data of the module $B_r$ are used for the entire period of the measurements (number of points: 33251). The approximation curve is shown by white color.

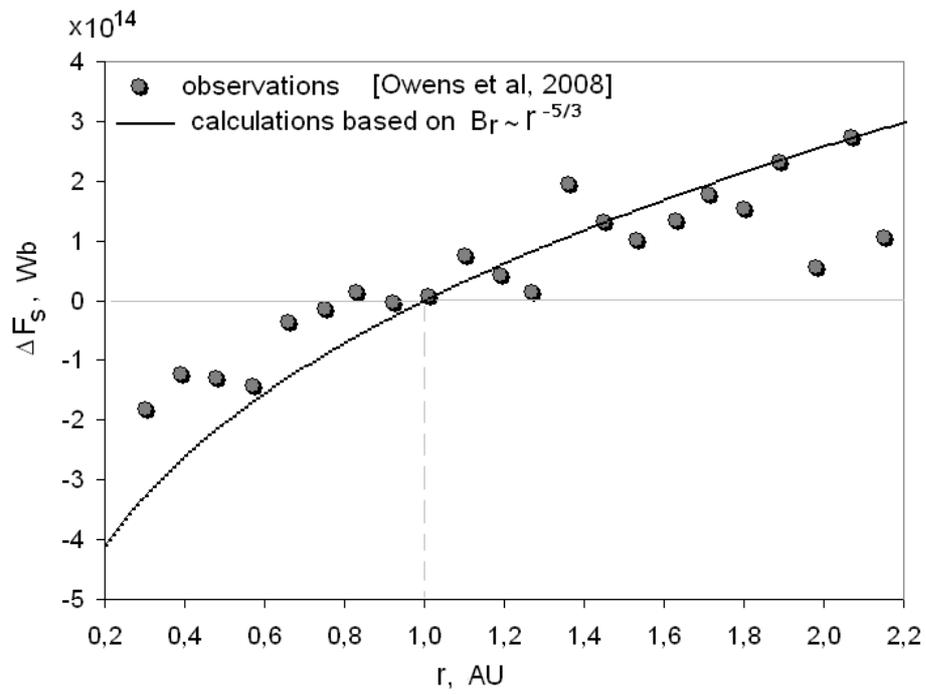

**Fig. 2.** Illustration of the cause of the "magnetic excess" effect. Points represent the difference between the magnetic flux according to distant spacecraft measurements and observations at the Earth's orbit, $\Delta F_S$ (from fig.5 in [8]). The curve is $\Delta F_S$ calculated on the basis of the formula (3), where the experimentally found dependence $B_r \propto r^{-5/3}$ is used.

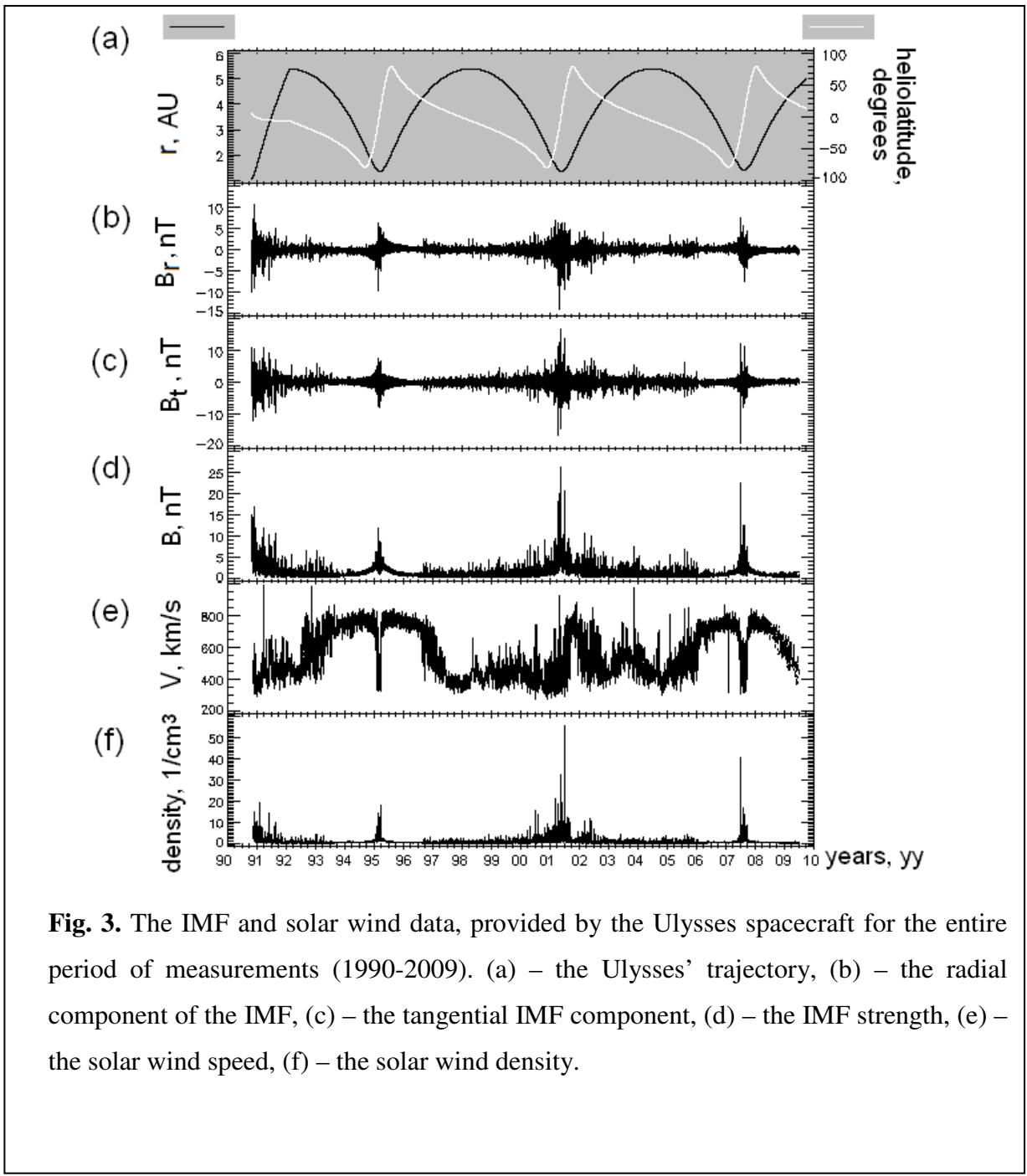

**Fig. 3.** The IMF and solar wind data, provided by the Ulysses spacecraft for the entire period of measurements (1990-2009). (a) – the Ulysses' trajectory, (b) – the radial component of the IMF, (c) – the tangential IMF component, (d) – the IMF strength, (e) – the solar wind speed, (f) – the solar wind density.

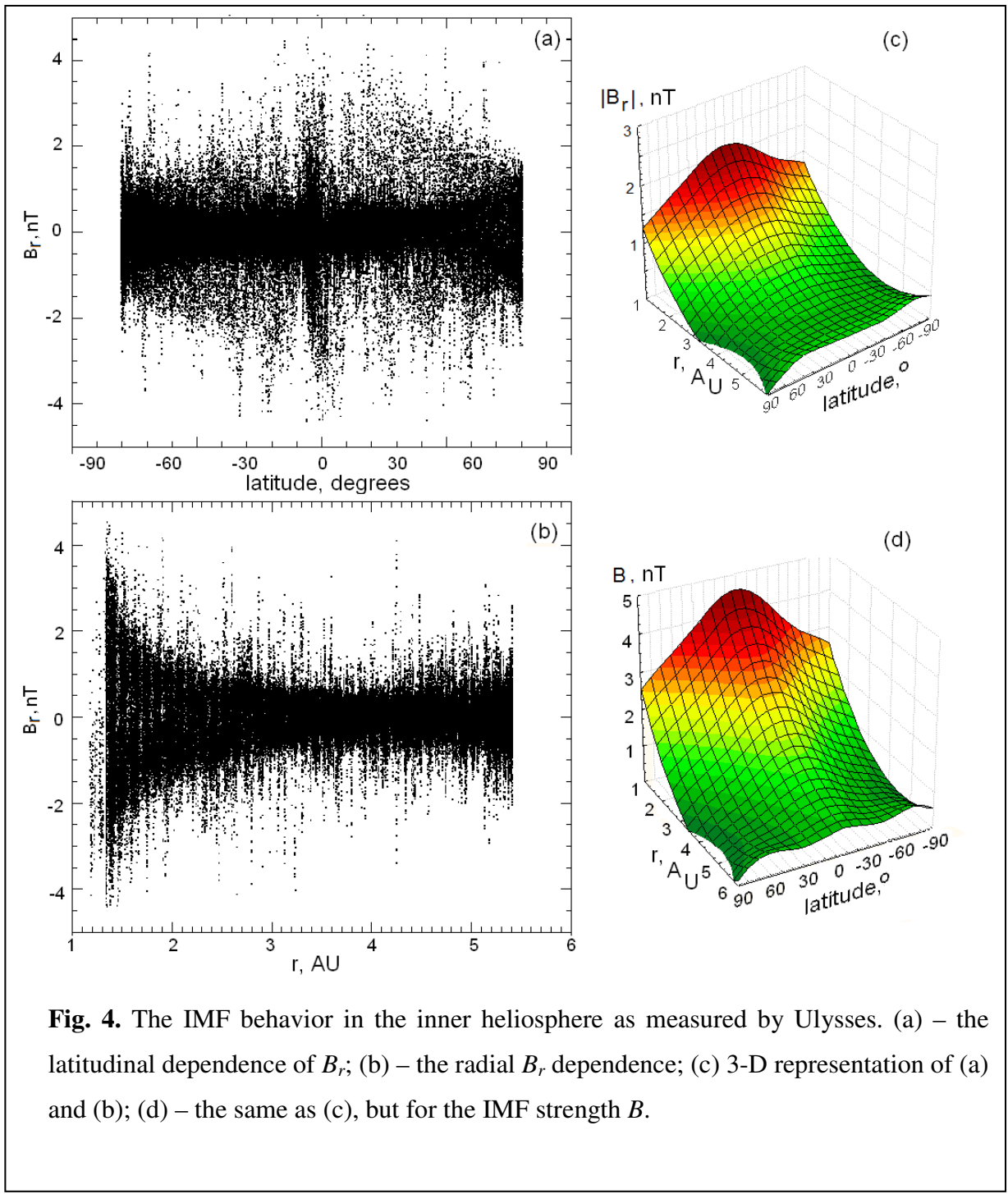

**Fig. 4.** The IMF behavior in the inner heliosphere as measured by Ulysses. (a) – the latitudinal dependence of $B_r$; (b) – the radial $B_r$ dependence; (c) 3-D representation of (a) and (b); (d) – the same as (c), but for the IMF strength $B$.

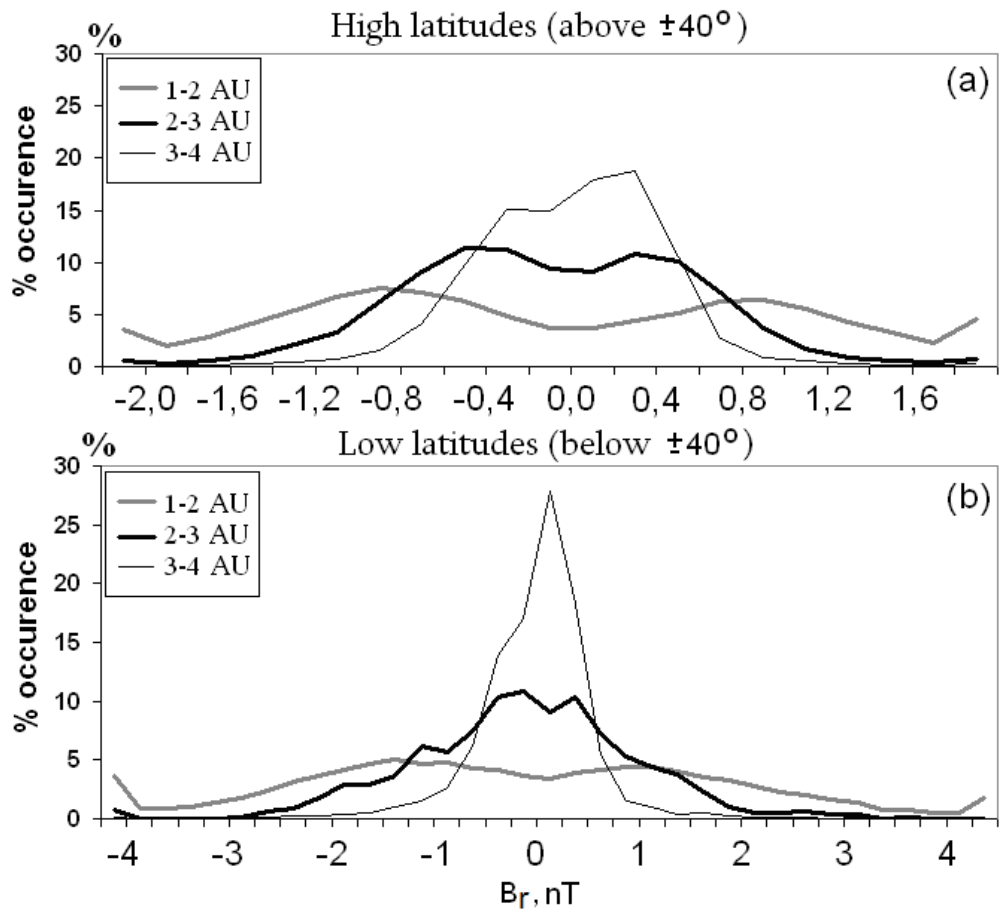

**Fig. 5.** The latitudinal dependence of the $B_r$ histogram seen at different heliocentric distances: (a) – high latitudes, (b) – low latitudes. The grey curve corresponds to 1-2 AU; the thick black curve – to 2-3 AU; the thin black curve – to 3-4 AU.

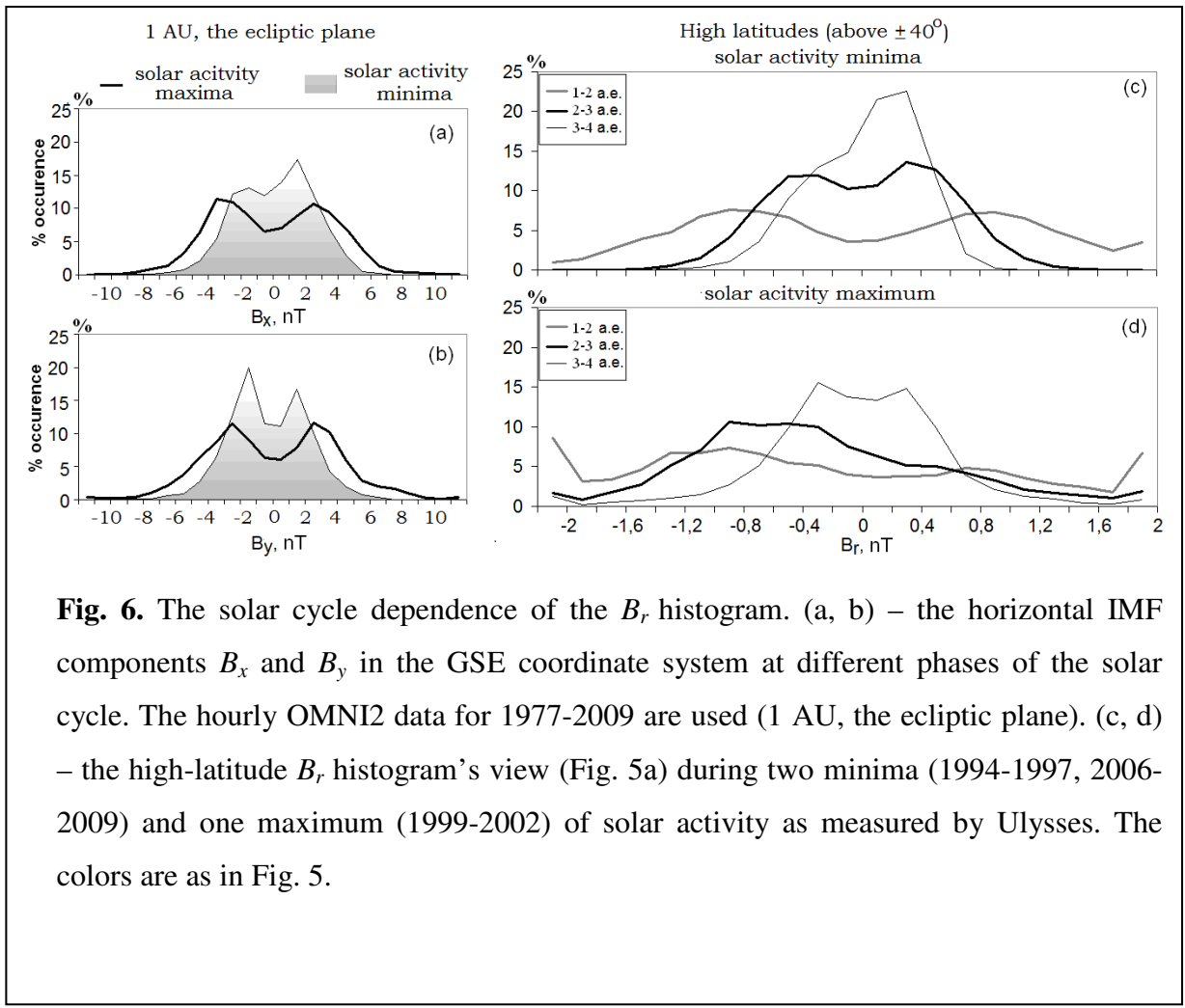

**Fig. 6.** The solar cycle dependence of the $B_r$ histogram. (a, b) – the horizontal IMF components $B_x$ and $B_y$ in the GSE coordinate system at different phases of the solar cycle. The hourly OMNI2 data for 1977-2009 are used (1 AU, the ecliptic plane). (c, d) – the high-latitude $B_r$ histogram's view (Fig. 5a) during two minima (1994-1997, 2006-2009) and one maximum (1999-2002) of solar activity as measured by Ulysses. The colors are as in Fig. 5.

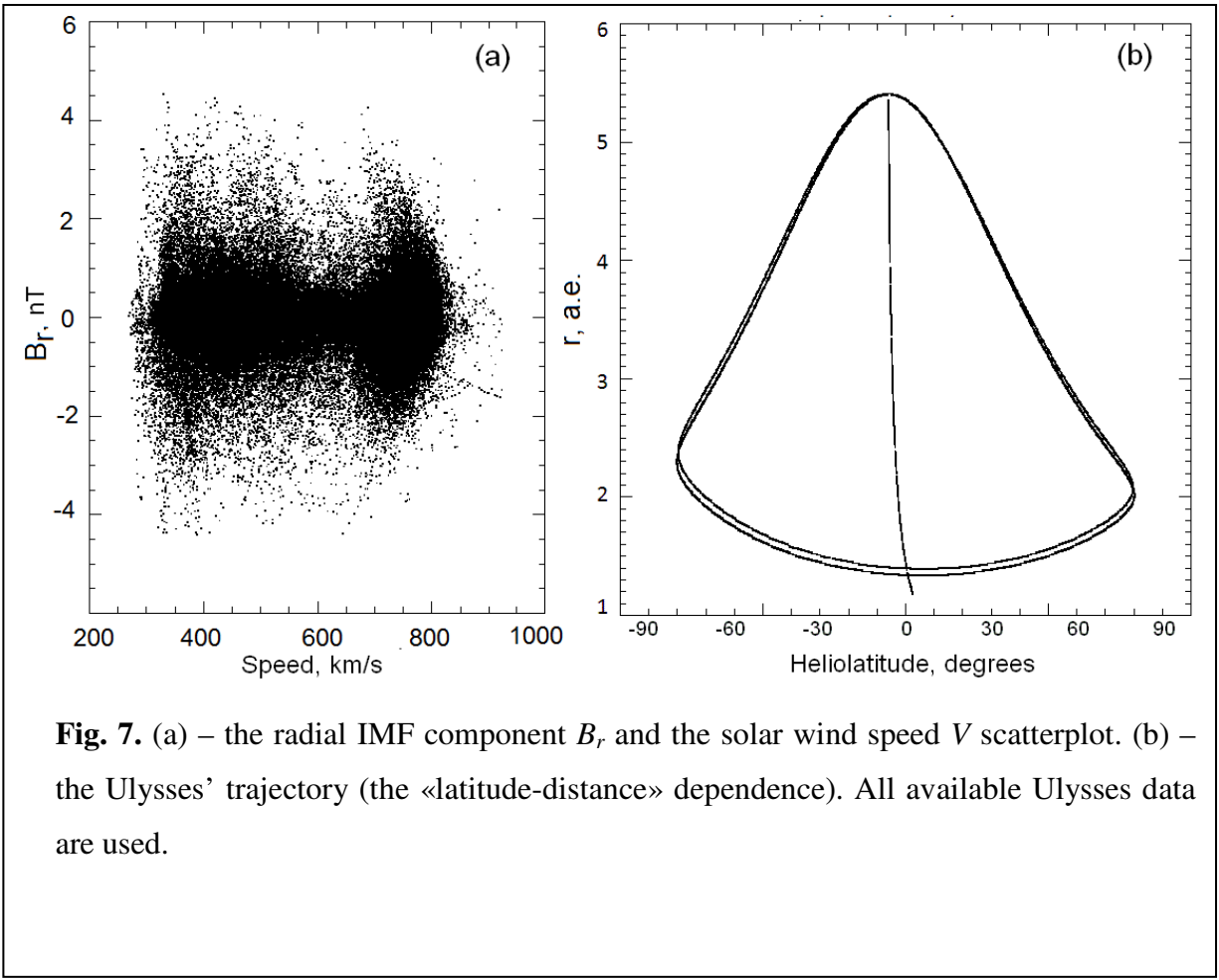

**Fig. 7.** (a) – the radial IMF component $B_r$ and the solar wind speed $V$ scatterplot. (b) – the Ulysses' trajectory (the «latitude-distance» dependence). All available Ulysses data are used.

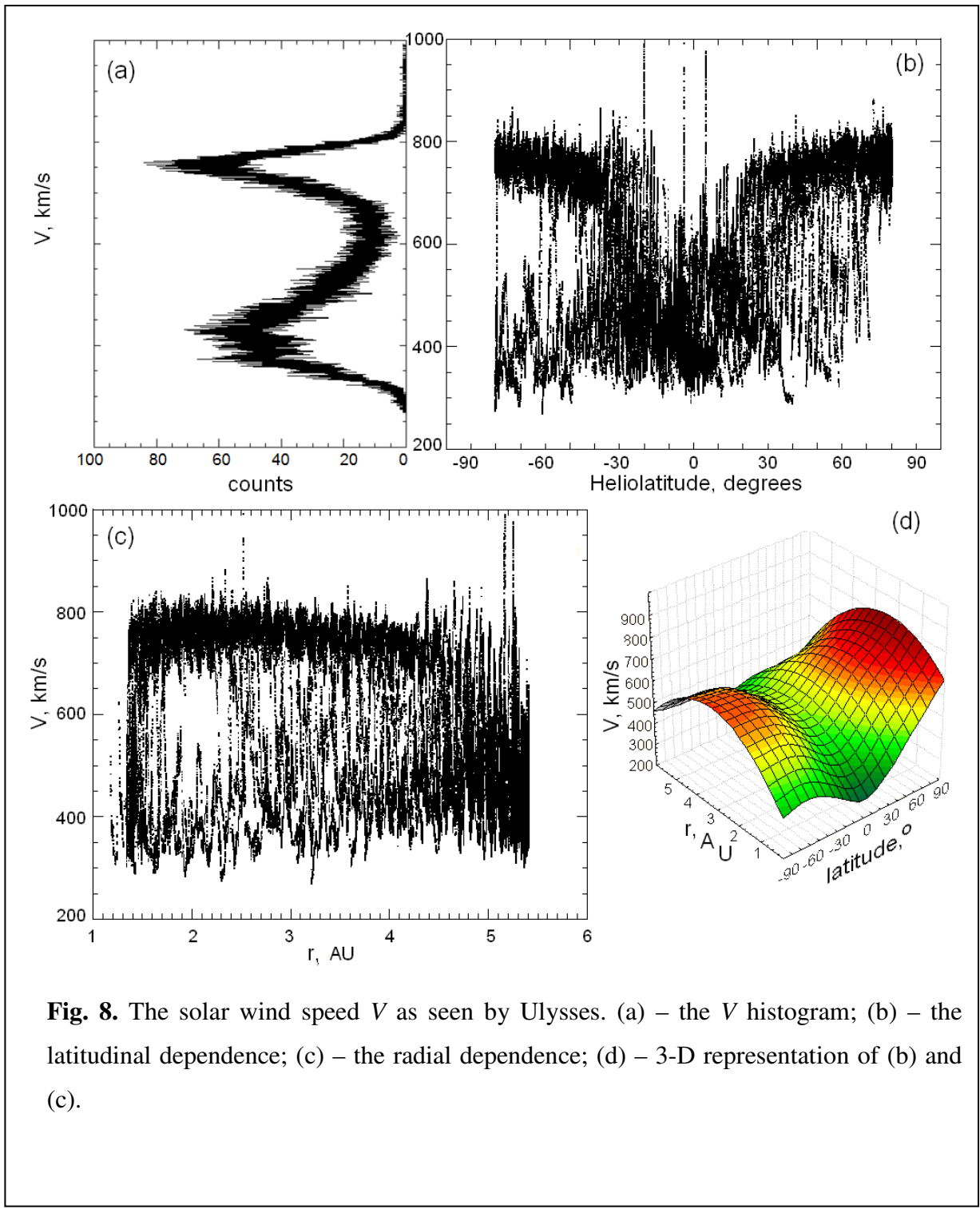

**Fig. 8.** The solar wind speed *V* as seen by Ulysses. (a) – the *V* histogram; (b) – the latitudinal dependence; (c) – the radial dependence; (d) – 3-D representation of (b) and (c).

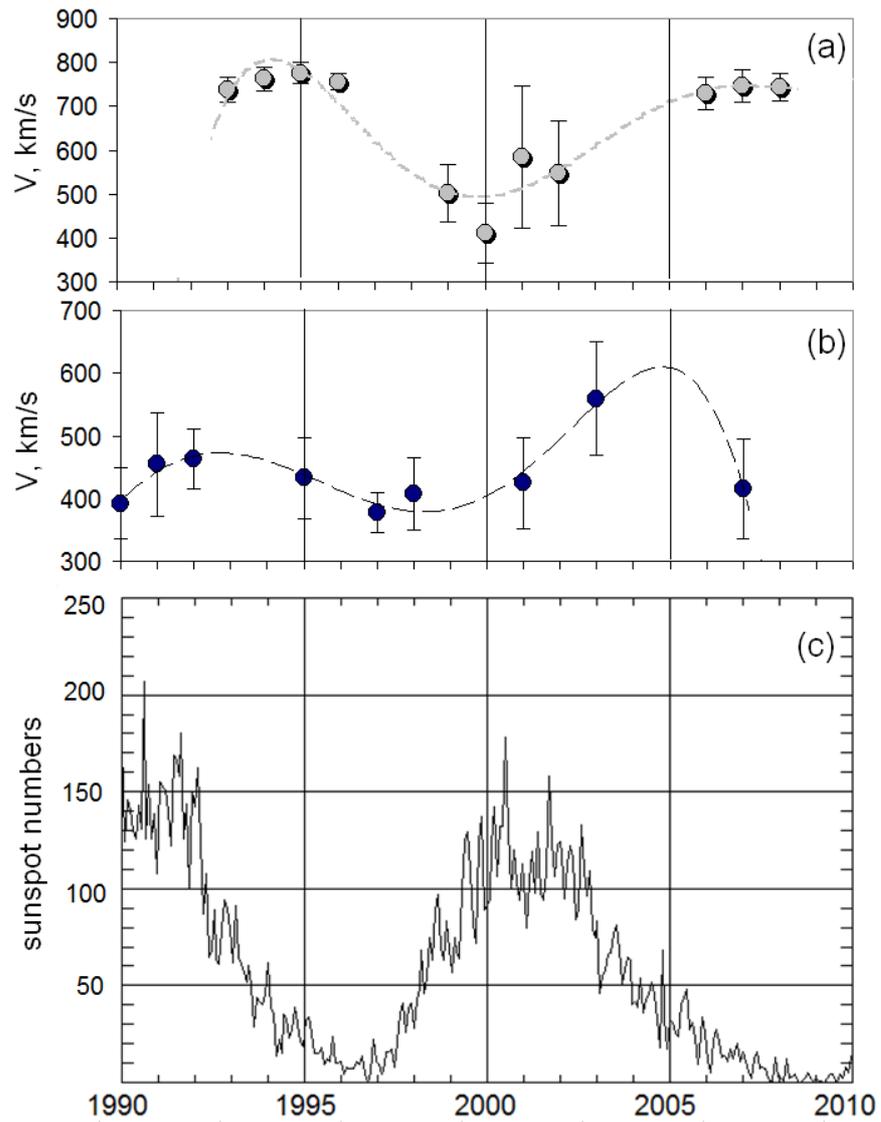

**Fig. 9.** The solar cycle dependence of the solar wind speed *V* on the basis of the Ulysses data. (a) – the annual mean of *V* at the heliolatitudes above 40°, (b) the annual mean of *V* near the ecliptic plane (±10° around), (c) the sunspot numbers (27-days averaging, the OMNI database).

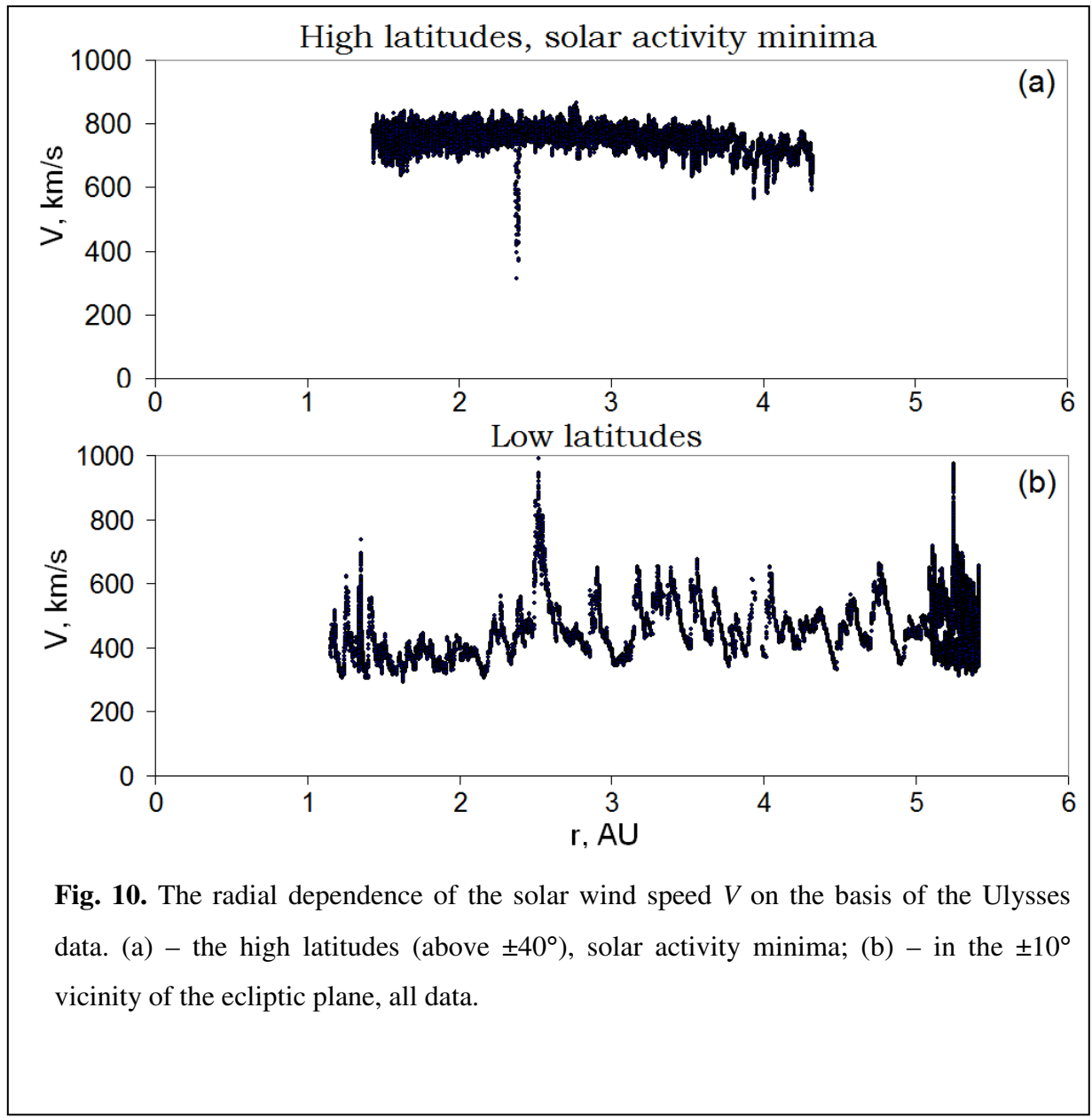

**Fig. 10.** The radial dependence of the solar wind speed *V* on the basis of the Ulysses data. (a) – the high latitudes (above ±40°), solar activity minima; (b) – in the ±10° vicinity of the ecliptic plane, all data.